\documentclass{article}
\usepackage[utf8]{inputenc}
\usepackage[T1]{fontenc}
\usepackage{authblk}
\usepackage{amssymb}
\usepackage{amsmath}
\usepackage{caption}
\usepackage{subcaption}
\usepackage{graphicx}
\usepackage{comment}
\usepackage[table,xcdraw]{xcolor}
\usepackage{xcolor,soul}
\definecolor{lightgreen}{rgb}{.7,95,.65}
\sethlcolor{lightgreen}

\title{Reduced Markovian descriptions of Brownian dynamics: toward an exact theory}

\author[1]{Matteo Colangeli\thanks{matteo.colangeli1@univaq.it}}
\author[2]{Adrian Muntean\thanks{adrian.muntean@kau.se}}
\affil[1]{Department of Information Engineering, Computer Science and Mathematics,
University of L'Aquila, Italy.}
\affil[2]{Department of Mathematics and Computer Science \& Centre for Societal Risk Research (CSR), Karlstad University, Sweden.}

\begin{document}

\maketitle
\begin{abstract}
We outline a reduction scheme for a class of Brownian dynamics which leads to meaningful corrections to the Smoluchowski equation in the overdamped regime. The mobility coefficient of the reduced dynamics is obtained by exploiting the Dynamic Invariance principle, whereas the diffusion coefficient fulfils the Fluctuation-Dissipation theorem. Explicit calculations are carried out in the case of a harmonically bound particle. A quantitative pointwise representation of the reduction error is also provided and
connections to both the Maximum Entropy method and the linear response theory are highlighted.
Our study paves the way to the development of reduction procedures applicable to a wider class of diffusion processes.
\end{abstract}

\section{Introduction}

The derivation of a contracted description of a Brownian particle subject to a confining potential is a long-standing problem in statistical mechanics, which dates back to an old question originally posed by Uhlenbeck and Ornstein \cite{OrUh}. For more background details and a general outline of the methods, we refer the reader to the seminal paper by van Kampen \cite{vanKam} as well as to the more recent reviews \cite{Givon,Legoll}. 
The Smoluchowski equation stands as a prominent example of a reduced description in the high friction regime, where the momentum variable rapidly thermalizes and the statistics of the particle is determined only by the distribution in the configuration space. In the last decades, a large research endeavour \cite{Wile,Shiba,Sancho} has pointed towards the derivation of corrections to the Smoluchowski formula for finite values of the friction constant. A classical iterative scheme, described e.g. in \cite{Risken}, derives solutions of the Kramers equation in terms of matrix continued fractions, via an expansion in powers of the inverse friction coefficient. Further guidelines on the derivation of the Smoluchowski equation from the Kramers equation can also be found in \cite{Boq,Cerr,Godd}.  A different approach, developed by Titulaer \cite{Titul78}, implements a Chapman-Enskog reduction scheme on the Fokker-Planck equation. A systematic use of the same procedure applied for the adiabatic elimination of fast variables was also considered in \cite{Theiss85}. More recently, a non-local version of the Smoluchowski equation was also obtained from the Kramers equation through the Chapman-Enskog procedure in \cite{Chav}. In the set-up of kinetic theory of gases, the Chapman-Enskog method has proved successful in the derivation of the Euler and the Navier-Stokes equations of fluid dynamics from the Boltzmann equation. However, as it was first demonstrated by Bobylev for Maxwell's molecules \cite{Bobyl06}, the Chapman-Enskog expansion is prone to small wavelength instabilities: namely, sufficiently short acoustic waves increase with time instead of decaying. This creates difficulties for an extension of hydrodynamics, as derived from a kinetic description, in the regime of finite Knudsen numbers, where the Navier-Stokes approximation is inapplicable. The study of various kinetic models \cite{colan07,colan07b,colan08,colan09} has later revealed that such instabilities can be cured by taking into account also the remote terms of the expansion. The resulting hydrodynamic equations, obtained from an exact summation of the Chapman-Enskog scheme, are indeed stable for all wavelengths, at variance with the finite-order approximations. The structure of the slow invariant manifold for systems displaying an intrinsic separation of time scales can be determined through a non-perturbative reduction procedure known as the invariant manifold method \cite{GorKar05}.
A key ingredient, in the method, is an equation of \textit{dynamic invariance} which, for a class of kinetic models known as linearized Grad's moment systems, was shown to lead to the same result as the exact summation of the Chapman-Enskog expansion \cite{Kar00,Kar02}. 
Thus, the plan of this work is to adopt, in the context of Brownian dynamics, the invariant manifold set-up to derive meaningful corrections to the Smoluchowski equation, while preserving the Markovian structure of the original process.
Unlike previous reduction schemes, the proposed procedure operates on the deterministic component of the dynamics, whereas the stochastic terms are properly handled via the Fluctuation-Dissipation theorem.
Explicit calculations can be carried out for the Brownian oscillator model, in which the equivalence between the condition of dynamic invariance and the exact summation of the Chapman-Enskog expansion is also established. The special case with a harmonic potential can hence be used as a test-bed to illustrate the general formalism.
With other models, instead, quasi-equilibrium manifolds typically constitute the starting point towards an iterative method of solution of the equation of invariance.

We envisage further developments of similar reduction schemes in the direction of coarse-graining of interacting particle systems as well as of partial differential equations with randomly fluctuating coefficients. Such research line might possibly connect this work with periodic and/or random homogenization questions; see \cite{Lions,Hong2,Luc-Pav,Hiroshi,Stuart} and  \cite{book_chapter} for recent applications of reduction schemes to statistical physics and epidemiological models. We also refer the reader to \cite{Kay18,stable_manifold,Roberts,Ruelle} for related matters, as well as to \cite{Lucarini,Daan} for applications of similar methods to the reduction of complex dynamics connected to climate change topics, where the need of developing innovative reduction techniques is growing. In this line of thinking, a rigorous characterization of slow invariant manifolds of random dynamical systems can be found in \cite{Ludw}. 
Other relevant questions related to the procedure of model reduction are also addressed in this work. One, for instance, concerns the derivation of a quantitative estimate of the reduction error stemming from the application of the method. The role of the initial data and of the \textit{defect of invariance} will be properly highlighted. Furthermore, as the contracted description retains just some of the observables of the original model, it is a non-trivial task to quantify to which extent a certain reduced dynamics yields a response to (small) perturbations comparable to that expressed by the original dynamics. The study of the linear response formalism will enable us to answer also this question.

This paper is organized as follows.
In Section \ref{sec:model}, we introduce the model and also illustrate the general set-up of the invariant manifold method and the related Chapman-Enskog scheme. In Sec. \ref{sec:results} we present some iterative schemes, based on the Maximum Entropy principle, which can be used to solve the invariance equation for a class of Brownian dynamics.  We focus, in particular, on the case of a harmonically bound Brownian particle, where the reduction procedure can be outlined in detail. We also provide a quantitative pointwise representation of the reduction error, and derive response functions due to the original and the reduced dynamics of the Brownian oscillator model. Finally, we draw our conclusions in Section \ref{sec:concl}.

\section{The model}
\label{sec:model}

In this work we exploit the Dynamic Invariance principle to derive reduced descriptions for a class of Brownian dynamics.
Specifically, we consider the Brownian dynamics of particles subject to a power law potential $V(x)=x^{n}$, $n\ge 2$ an integer, described by a system of stochastic differential equations (SDEs) written in the It\^{o} form:
\begin{eqnarray}
dx(t)&=&v(t) dt \nonumber\\
dv(t)&=& -\frac{1}{m}\partial_xV(x)dt- \gamma v(t) dt +\sqrt{2 D \gamma^2} dW(t) \; , \label{underd}
\end{eqnarray}
where $W(t)$ is a one-dimensional Wiener process, $m$ is the mass, $\gamma$ is the friction constant, $D=(\beta m \gamma)^{-1}$ is the diffusion coefficient, and $\beta$ is the inverse temperature of the system.
We recall that the leading high friction approximation of the set of equations \eqref{underd} is commonly written in the Smoluchowski form:
\begin{equation}
dx(t)=-\frac{1}{m\gamma}\partial_xV(x)dt+\sqrt{2 D} dW(t) \; . \label{overd3}
\end{equation}
We hence seek for a reduced description resembling the structure of Eq. \eqref{overd3} and based on the following SDE:
\begin{equation}
dx(t)=-\chi  \partial_xV(x) dt+\sqrt{2 \mathcal{D}_r}dW(t) \; , \label{nonlinLE}
\end{equation}
where $\chi$ is the bare mobility and $\mathcal{D}_r$ the diffusion coefficient of the reduced dynamics. The two coefficients $\chi$ and $\mathcal{D}_r$ are expected to reduce to $(m\gamma)^{-1}$ and $D$, respectively, in the overdamped limit expressed by Eq. \eqref{overd3}. The derivation of the mobility $\chi$, in Eq. \eqref{nonlinLE}, will be tackled in Sec. \ref{sec:results} by exploiting the Dynamic Invariance principle, which is shortly reviewed next.

\subsection{The Dynamic Invariance principle}
\label{sec:DIP}
The invariant manifold method is a procedure of model reduction that was originally introduced as a special analytical perturbation technique in the KAM theory of integrable Hamiltonian systems \cite{Kol,Arn,Mos}. 
The method was later exploited in the kinetic theory of gases to derive the evolution equations of the hydrodynamic fields from the Boltzmann equation or related kinetic models \cite{Kar00,Kar02,colan08}. 
The basic picture underlying the invariant manifold method can be shortly summarized as follows, see Refs. \cite{GorKar05,GorKar}. There exists a manifold of slow motions, in the phase space of the system, parameterized by a set of distinguished macroscopic variables, which is \textit{positively invariant}: if a trajectory starts on the manifold at time $t_0$, it will remain on the manifold for all times $t>t_0$. Trajectories starting from arbitrary initial conditions quickly reach a neighborhood of the manifold, and then evolve along such slow manifold, until the equilibrium state is eventually attained. More explicitly, for kinetic equations we let $f$ denote the (single-particle) distribution function, whose evolution in a domain $U$ is described by the equation:
\begin{equation}
\partial_t f=J(f) \label{kin} \; .
\end{equation}
Let $m:f\rightarrow M$ be a surjective linear map, with $M$ denoting the macroscopic variables (moment fields) and also let $f_M$ denote a manifold
parametrized by a set of macroscopic fields $M$. We look for an invariant manifold $f_M$ parametrized by the value of $M$, obeying the self-consistency condition $m(f_M ) = M$.
The ``microscopic'' and ``macroscopic''  time derivatives of $f$ on the manifold $f_M$ are defined as:
\begin{eqnarray}
\partial_t^{(micro)}f_M&:=&J(f_M)  \label{dtmicro} \; , \\
\partial_t^{(macro)}f_M&:=&\left(D_M f_M  \right) m(J(f_M))) \label{dtmacro} \; ,
\end{eqnarray}
where the differential $D_M f_M$, in Eq. \eqref{dtmacro}, is evaluated at the point $M=m(f_M)$. 
While $J(f_M)$ in Eq. \eqref{dtmicro}, corresponds to a value of the vector field $J$ evaluated on the manifold $f_M$, Eq. \eqref{dtmacro} codifies a chain rule: one computes the time derivative of the moment $M$ via the map $m$, as $\dot{M}=m(J(f_M))$, and then exploits the time dependence of $f$ which is expressed through the time dependence of $M$.
The invariance equation is written in the form:
\begin{equation}
\partial_t^{(micro)}f_M=\partial_t^{(macro)}f_M \; , \label{InvEq1}
\end{equation}
or, alternatively, as:
\begin{equation}
\Delta_M:=\partial_t^{(micro)}f_M-\partial_t^{(macro)}f_M=0 \; , \label{InvEq1}
\end{equation}
where $\Delta_M$ is called \textit{defect of invariance}.
The Dynamic Invariance principle requires that the equality in Eq. \eqref{InvEq1} is fulfilled for any values of the macroscopic variables $M$.
Solutions to Eq. \eqref{InvEq1} have been obtained from the study of various kinetic models, see e.g. \cite{colan07,colan07b,colan09}. Determining the structure of $f_M$ constitutes an instance of the \textit{closure problem} in kinetic theory \cite{Cer}. 
Note, indeed, that if a solution of Eq. \eqref{InvEq1} is found, then the moments $M$ obey the following closed system of evolution equations:
\begin{equation}
\dot{M}=m(J(f_M)) \label{hydro}\; .
\end{equation} 
We also recall that if a functional $E(f)$ is conserved for the dynamics \eqref{kin}, then $E(f_M)$ remains constant along the trajectories of the reduced system described by Eq. \eqref{hydro}. Moreover, if the time derivative of a functional $H(f)$ is nonpositive due to the dynamics \eqref{kin}, then so is also the time derivative of the functional $H(f_M)$ due to the reduced dynamics \eqref{hydro}.

Appropriate iterative schemes have been developed to solve the invariance equation \eqref{InvEq1}. One of the first systematic procedures of constructing invariant manifolds was the celebrated Chapman-Enskog  method for the Boltzmann equation \cite{CC}, whose main steps are also shortly recalled here. 
One starts with the singularly perturbed dynamics:
\begin{equation}
\partial_t f+A(f)=\frac{1}{\varepsilon} Q(f) \label{pert} \; ,
\end{equation}
with $\varepsilon>0$ a small parameter.
One requires that $m(Q(f))=0$ and that, for each $M\in m(U)$, the equation $Q(f)=0$, with $m(f)=M$, has a unique solution denoted by $f_M^{eq}$ (corresponding to the Maxwellian distributions, in Boltzmann's theory), which is also asymptotically stable and globally attracting for the fast dynamics 
\begin{equation} 
\partial_t f=\frac{1}{\varepsilon} Q(f) \; .
\end{equation}
The invariance equation \eqref{InvEq1} can be adapted to the singularly perturbed system \eqref{pert} in the form:
\begin{equation}
\frac{1}{\varepsilon} Q(f_M)=A(f_M)+(D_M f_M)(m(A(f_M))) \; .\label{InvEq3}
\end{equation}
In the Chapman-Enskog scheme, a solution to Eq. \eqref{InvEq3} is sought in the form of a series in powers of a small parameter $\varepsilon$:
\begin{equation}
f_M=f_M^{eq}+\sum_{i=1}^{\infty} \varepsilon^i f_M^{(i)} \; .
\end{equation}
In the set-up of Boltzmann's theory, the zero-order approximation $f_M\simeq f_M^{eq}$ leads, upon integrating the Boltzmann equation over the velocity space, to the inviscid Euler equations of fluid dynamics, whereas the first-order correction $f_M\simeq f_M^{(eq)}+\varepsilon f_M^{(1)}$ gives rise to the compressible Navier-Stokes equations supplied with transport coefficients which depend on the underlying collision model.

 \section{Results}
 \label{sec:results}

The aim of this Section is to exploit the method of the invariant manifold to derive suitable expressions for the mobility $\chi$ introduced in Eq. \eqref{overd3}. In Sec. \ref{sec:quasieq} we thus address the general case described by Eq. \eqref{underd}, and outline useful iterative schemes based on the Maximum Entropy principle. The Brownian oscillator model, corresponding to the case $n=2$, is studied in detail in Sec. \ref{sec:BO}. Next, the analysis of the reduction error and of correlation functions is deferred to Secs. \ref{sec:estim} and \ref{sec:LRT}.
 
\subsection{Quasi-equilibrium closures}
\label{sec:quasieq}

We start with the Fokker-Planck equation associated to the SDE \eqref{nonlinLE}, which reads:
\begin{equation}
\frac{\partial \rho(x,t)}{\partial t}=\mathcal{D}_r \frac{\partial}{\partial x} \left(\rho(x,t) \partial_x U(x) +  \frac{\partial}{\partial x}\rho(x,t)\right)\label{FPred2} \; ,
\end{equation}
where we enforced the Einstein relation $\chi=\beta \mathcal{D}_r$ and set $U(x)=\beta V(x)$. 
Letting $\mathbf{x}(t)=(x(t),v(t))$, we denote by $\langle \dots\rangle$ the conditional average referring to a deterministic initial datum $\mathbf{x}(0)=\mathbf{x}=(x,v)$.
We then write $M(t)\equiv \langle \partial_xU(x)\rangle$ and introduce the closure:
\begin{equation}
\langle v(t)\rangle=-\chi M(t) \label{closNL} \; ,
\end{equation}
where $M(t)$ plays the role of a slow variable, driving the evolution of the fast variable $\langle v(t)\rangle$. Application of the Dynamic Invariance principle, introduced in Sec. \ref{sec:DIP}, requires the evaluation of the two operators $\partial_t^{(micro)} \langle v(t)\rangle$ and $\partial_t^{(macro)} \langle v(t)\rangle$ for the dynamics obtained upon averaging Eqs. \eqref{underd} over noise. We thus find:
\begin{eqnarray}
\partial_t^{(micro)} \langle v(t)\rangle &:=&-\frac{1}{m}M(t)+\gamma \chi M(t) \label{ceNL1}\\
\partial_t^{(macro)} \langle v(t)\rangle &:=&-\chi \dot{M}(t) \label{ceNL2}
\end{eqnarray}
Upon equalizing the two expressions in Eqs. \eqref{ceNL1} and \eqref{ceNL2} one establishes the invariance equation
\begin{equation}
-\frac{1}{m}M(t)+\gamma \chi M(t)+\chi \dot{M}(t)=0 \; , \label{IE2}
\end{equation}
to be solved for the mobility $\chi$. Note that the classical form $1/(m \gamma)$  of the mobility is recovered from Eq. \eqref{IE2} as $\gamma\rightarrow \infty$, $m\rightarrow 0$ with $\gamma m$ finite (provided $\chi$ remains bounded as well). In order to solve Eq. \eqref{IE2}, and find corrections to the Smoluchowski equation, it is thus necessary to determine an explicit expression for $\dot{M}(t)$. This task can be pursued iteratively via the Maximum Entropy method (MaxEnt), often invoked in statistical mechanics; see e.g. \cite{VulRon17,GorKar05}. We start by defining the entropy:
\begin{equation}
S[\rho]=-\int \rho(x,t) \ln \frac{\rho(x,t)}{\rho_0(x)}dx \; , \label{entropy}
\end{equation}
which is a monotonically growing function attaining its maximum at equilibrium, i.e. when $\rho=\rho_0$. 
Let $M=\{M_0,\dots,M_k\}$ be a set of linearly independent moments of $\rho$ defined as:
\begin{equation}
\int m_i(x) \rho(x,t)dx=M_i(t) \quad,\quad i=0,\dots,k \;, \label{maxent}
\end{equation}
where the $m_i$'s are the microscopic densities of the moments, with $m_0=1$. The so-called \textit{quasi-equilibrium} density $\rho^*(x,M)$ is obtained by maximising the entropy $S[\rho]$ under the constraints of fixed $M$, which yields:
\begin{equation}
\rho^*(x,M)=\rho_0(x)\exp\left[\sum_{i=0}^k \Lambda_i m_i(x)\right] \label{qe}
\end{equation}
where $\Lambda=\{\Lambda_1,\dots,\Lambda_k\}$ denotes the set of Lagrange multipliers which depend on the set of moments $M$. The quasi-equilibrium projector is then defined as:
\begin{equation}
\Pi^* \bullet = \sum_{i=0}^k \frac{\partial \rho^*}{\partial M_i}\int m_i(x) \bullet dx  \label{proj}
\end{equation}
By acting with the projector $\Pi^*$ on both sides of Eq. \eqref{FPred2} one obtains the following moment equations in the quasi-equilibrium approximation:
\begin{equation}
\dot{M}_i=-\sum_{j=0}^k \Lambda_j \int \partial_x m_i(x) \rho^*(x,t) \partial_x m_j(x) dx \; . \label{QE1}
\end{equation} 
A step forward can be made by splitting the set of moments as $M=M_I \cup M_{II}$, with $M_I=\{M_0,\dots,M_\ell\}$ and $M_{II}=\{M_{\ell+1},\dots,M_k\}$. We assume that the (first) quasi-equilibrium distribution $\rho^*(x)$ can be derived explicitly for the set $M_I$, i.e. $\rho^*=\rho^*(x,M_I)$, and we hence seek for the second quasi-equilibrium closure in the form $\rho=\rho^*(1+\varphi)$. An expansion of the functional \eqref{entropy} in a neighbourhood of $\rho^*$ to second order in $\varphi$ yields:
\begin{equation}
\Delta S[\varphi]=-\int \rho^* \varphi \left[1+\ln \frac{\rho^*}{\rho_0}  \right]dx-\frac{1}{2}\int \rho^* \varphi^2 dx \; . \label{entropy2}
\end{equation}
The deviation $\varphi$ is determined from the following maximization problem, called ``Triangle MaxEnt approximation'' \cite{Karlin16}:
\begin{equation}
\Delta S \rightarrow \textrm{max} \; , \; \int m_I(x) \rho^* \varphi\; dx =0 \; , \; \int m_{II}(x) \rho^* \varphi\; dx =\Delta M_{II} \; ,
\end{equation}
where $\Delta M_{II}\equiv M_{II}-M_{II}(M_I)$ represents the deviations of the moments $M_{II}$ from the values attained in the first quasi-equilibrium state.

To solve the invariance equation \eqref{IE2}, we restrict to the one-moment quasi-equilibrium closures, i.e. we set $M_I=M_0$ (i.e. $\rho^*(x)=\rho_0(x)$) and $M_{II}=M$, with $\int m(x) \rho\, dx = M$.
We thus find:
\begin{equation}
\frac{\partial \varphi}{\partial t}=\mathcal{L}\varphi, \quad , \quad \mathcal{L}\bullet = \rho_0^{-1}\mathcal{D}_r \frac{\partial}{\partial x}  \rho_0 \frac{\partial}{\partial x} \bullet \; . \label{evol}
\end{equation}
 The triangle one-moment quasi-equilibrium distribution is found in the form:
 \begin{equation}
 \varphi^{(0)}(x,t)=m^{(0)}(x)\left(M(t)-\langle M \rangle_0\right) \label{triangle} \; ,
 \end{equation}
with 
\begin{equation}
m^{(0)}(x)=\frac{m(x)-\langle m \rangle_0}{\langle  m^2\rangle_0-\langle  m\rangle_0^2} \label{m0}  \;.
\end{equation}
We can thus construct a refinement of the quasi-equilibrium dynamics \eqref{QE1}, by defining the new projection operator:
\begin{equation}
\Pi^{(0)} \bullet = \rho_0(x) \frac{m^{(0)}(x)}{\langle  m^{(0)} m^{(0)}\rangle_0}\int m^{(0)}(x) \bullet dx  \label{proj2} \; .
\end{equation}
After inserting the expression \eqref{triangle} in \eqref{evol} and upon acting on both sides with $\Pi^{(0)}$, one finally obtains:
\begin{equation}
\dot{M}=-\ell_0 \left(M(t)-\langle M \rangle_0\right)  \label{macro} \; ,
\end{equation}
with
\begin{equation}
\ell_0=-\frac{\langle  \partial_x m^{(0)}\mathcal{D}_r \partial_x m^{(0)}  \rangle_0}{\langle  m^{(0)} m^{(0)} \rangle_0} \label{elle0} \; ,
\end{equation}
which defines the relaxation dynamics of the moment $M$ to the corresponding value attained in the equilibrium state.
We observe that the strategy exploited so far can be prosecuted further, by using the iterative scheme discussed in Ref. \cite{Karlin16}:
\begin{equation}
(1-\Pi^{(j)}) \mathcal{L}m^{(j+1)}=0 \; , \label{iter}
\end{equation}
with $j=0,1,\dots$, where $m^{(j+1)}=m^{(j}+\mu^{(j+1)}$, and in which the orthogonality condition $\langle \mu^{(j+1)} m^{(j}\rangle_0=0$ is exploited.
At the k-th iteration step, a refinement of the inverse relaxation time \eqref{elle0} attains the form:
\begin{equation}
\ell_k=-\frac{\langle  \partial_x m^{(k)}\mathcal{D}_r \partial_x m^{(k)}  \rangle_0}{\langle  m^{(k)} m^{(k)} \rangle_0} \label{ellek} \; .
\end{equation}
The sequence $(\ell_k)_{k\in \mathbb{N}}$ is found to converge to the eigenvalue of the operator $\mathcal{L}$ with the  minimal non-zero absolute value. 

In the next Section we will address the case with $n=2$, corresponding to a harmonically bound particle, which can be solved explicitly without adopting the foregoing iterative scheme based on the MaxEnt principle. 

\subsection{The Brownian oscillator}
\label{sec:BO}
We now turn to study in detail the case of a Brownian particle bounded in a harmonic potential $V(x)=1/2  m \omega_0^2 x^2$, which constitutes one of the classical exactly solvable models of nonequilibrium statistical mechanics \cite{chandra}. We refer also the reader to Ref. \cite{Metzler} for some recent inspection of the ergodic properties of Ornstein-Uhlenbeck systems.
The dynamics of the Brownian oscillator is described by a system of linear SDEs:
\begin{eqnarray}
dx(t)&=&v(t) dt \nonumber\\
dv(t)&=& -\omega_0^2 x(t) dt- \gamma v(t) dt +\sqrt{2 D \gamma^2} dW(t) \; , \label{eq1}
\end{eqnarray}
where $\omega_0=\sqrt{k/m}$ is the natural frequency of the  oscillator with mass $m$ and elastic constant $k$, see Appendix \ref{app:appA} for details.  We shall refer to Eq. \eqref{eq1} as the \textit{original} dynamics of the Brownian oscillator. 
We introduce the \textit{drift} matrix $\mathbf{M}$, defined as:
\begin{equation}
    \mathbf{M}=\begin{pmatrix}
0 & -1 \\
\omega_0^2 & \gamma
\end{pmatrix} \, , \label{M}
\end{equation}
 whose eigenvalues read:
\begin{equation}
    \lambda_{\pm}=\frac{\gamma\pm \gamma_s}{2}, \label{eigenv}
\end{equation}
with $\gamma_s=\sqrt{\gamma^2-4 \omega_0^2}=\lambda_+-\lambda-$. 
Henceforth we shall restrict our discussion to the overdamped regime, namely the region in the parameter space in which $\gamma_s$ is real and larger than zero.\\
An exact reduced description, not requiring a separation of time scales, is available for the Brownian oscillator model \cite{Zwanzig}.
This is obtained by integrating over time the second of Eqs. \eqref{eq1} and by then inserting the obtained expression in the first equation. The resulting reduced dynamics, expressed in terms of the configuration variable $x(t)$, turns out to be non-Markovian. Nevertheless, in the regime of high friction, and for times much longer than $\gamma^{-1}$, the Markovian structure of the reduced dynamics can be restored \cite{Sancho,Luczka}. \\
Another contracted description of the model can instead be derived by considering the overdamped limit of Eq. \eqref{eq1} \cite{Pavl}, that is worth briefly recalling.
By letting $\mathbf{x}^{\varepsilon}(t)=\mathbf{x}(\varepsilon^{-1}t)$, with $\varepsilon=\gamma^{-1}$, the original dynamics can be rescaled as follows:
\begin{eqnarray}
    dx^{\varepsilon}(t)&=& \varepsilon^{-1} v\ dt \label{pv1}\\
   dv^{\varepsilon}(t) &=&- \varepsilon^{-1}\omega_0^2 x^{\varepsilon}(t)\ dt - \varepsilon^{-2} v^{\varepsilon}(t)\ dt+\varepsilon^{-1}\sqrt{2 (\beta m)^{-1}} dW(t), \label{pv2} 
   \end{eqnarray}
where we exploited the scaling $dW(\epsilon^{-1}t)=\epsilon^{-1/2}dW(t)$.
It thus holds:
\begin{equation}
    \varepsilon^{-1}v^{\varepsilon}(t)dt=-\omega_0^2 x^{\varepsilon}(t)\ dt+\sqrt{2 (\beta m)^{-1}} dW(t)+\mathcal{O}(\varepsilon), \label{pv3}
\end{equation}
and hence,
\begin{equation}
    dx^{\varepsilon}(t)=-\omega_0^2 x^{\varepsilon}(t)\ dt+\sqrt{2 (\beta m)^{-1}} dW(t)+\mathcal{O}(\varepsilon). \label{pv4}
\end{equation}
As $\epsilon\rightarrow 0$, Eq. \eqref{pv4} leads to the Smoluchowski equation for the Brownian oscillator, which, after turning back to the original variables, attains the well-known structure:
\begin{equation}
dx(t)=-\frac{\omega_0^2}{\gamma}x(t) dt+\sqrt{2 D}\,dW(t) \, . \label{smoluc}
\end{equation}
Let us now turn to illustrate our reduction scheme, based on the Dynamic Invariance principle.
We aim at setting up a reduced description which formally resembles the structure of Eq. \eqref{smoluc}, and is based on the linear SDE:
\begin{equation}
dx(t)=-\alpha x(t) dt+ \sqrt{2 \mathcal{D}_r} \, dW(t)  \, ,\label{red}
\end{equation}
where $\alpha$ and $\mathcal{D}_r$ denote the drift and diffusion coefficients of the \textit{reduced} dynamics, respectively. These coefficients will properly recover the two corresponding expressions $\omega_0^2/\gamma$ and $D$ attained in the overdamped limit, cf. Eq. \eqref{smoluc}.
We remark that unlike the inverse friction expansion method, see e.g. \cite{Risken}, which typically starts from the Kramers equation associated to Eq. \eqref{eq1} and yields Eq. \eqref{smoluc} as the leading-order term of an expansion in powers of $\varepsilon=\gamma^{-1}$, our scheme applies to the deterministic component of the Brownian dynamics. The drift coefficient $\alpha$ will be obtained from the solution of an invariance equation, which is derived either from an exact summation of the Chapman-Enskog expansion or, equivalently, by the application of the Dynamic Invariance principle. The coefficient $\mathcal{D}_r$ is then found by exploiting the Fluctuation-Dissipation theorem. In the sequel we shall address, separately, the two distinct steps of the reduction procedure. 

\begin{figure}[h!]
\begin{center}
         \includegraphics[width=0.32\textwidth]{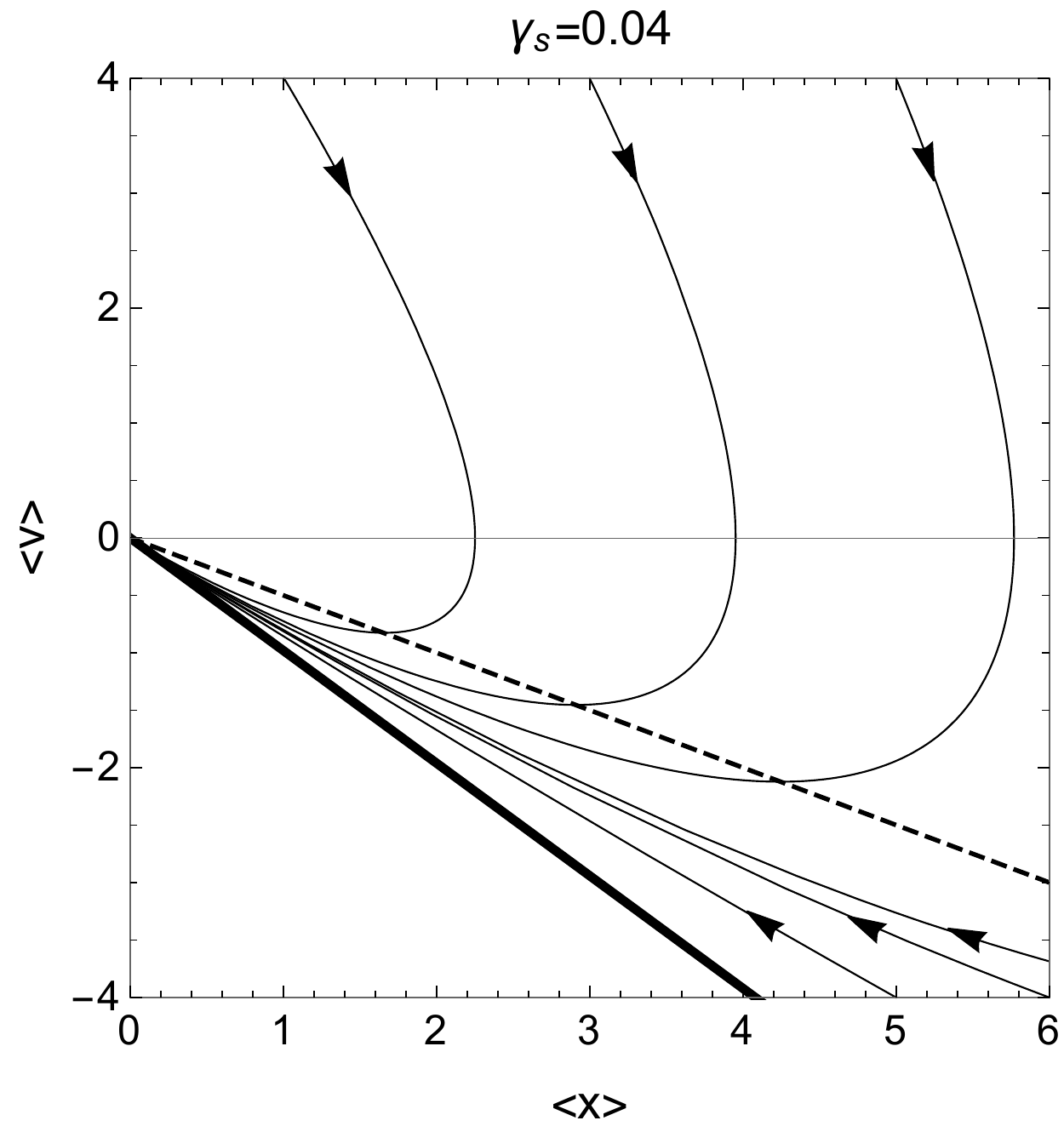}
         \includegraphics[width=0.32\textwidth]{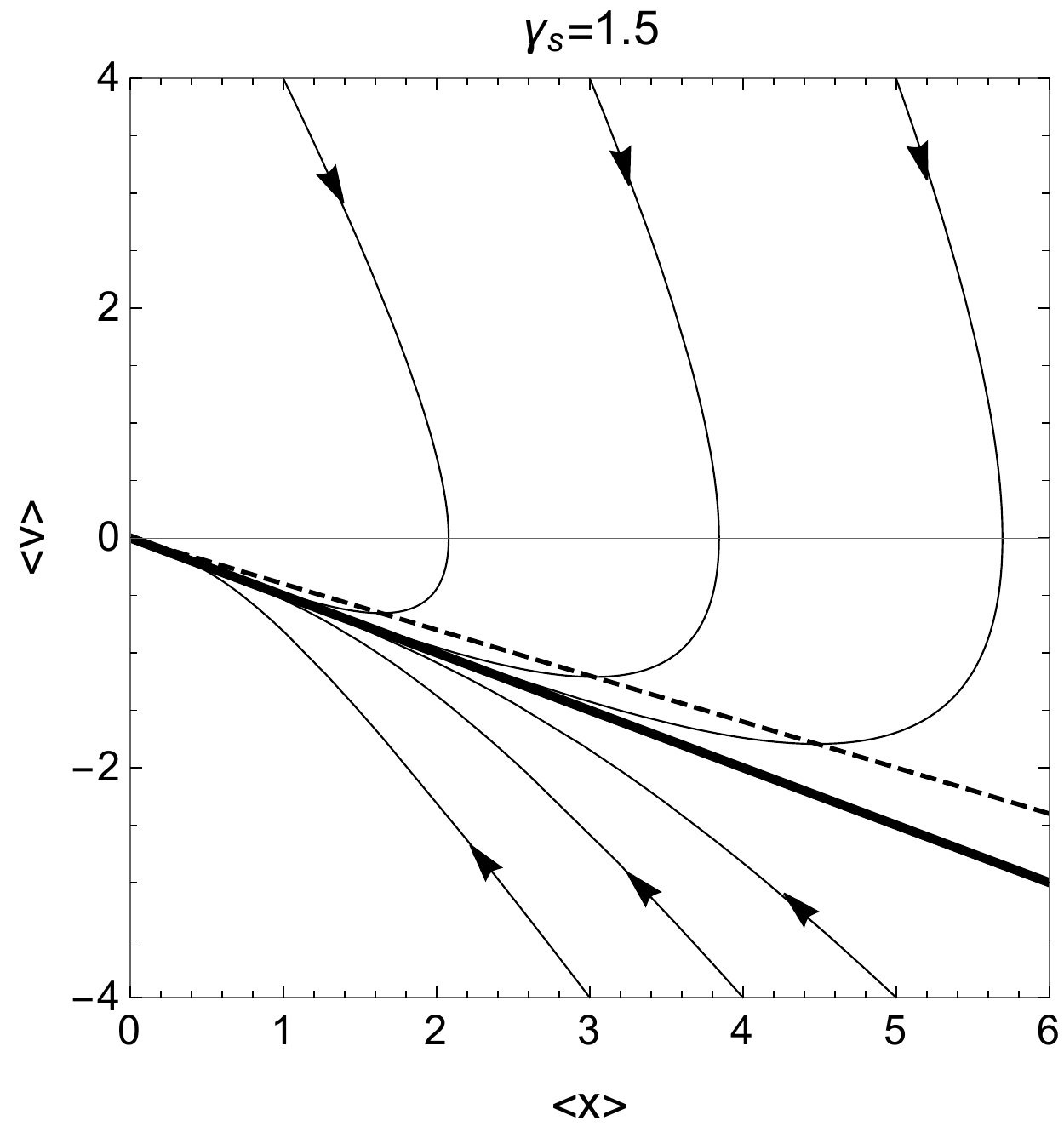}
         \includegraphics[width=0.32\textwidth]{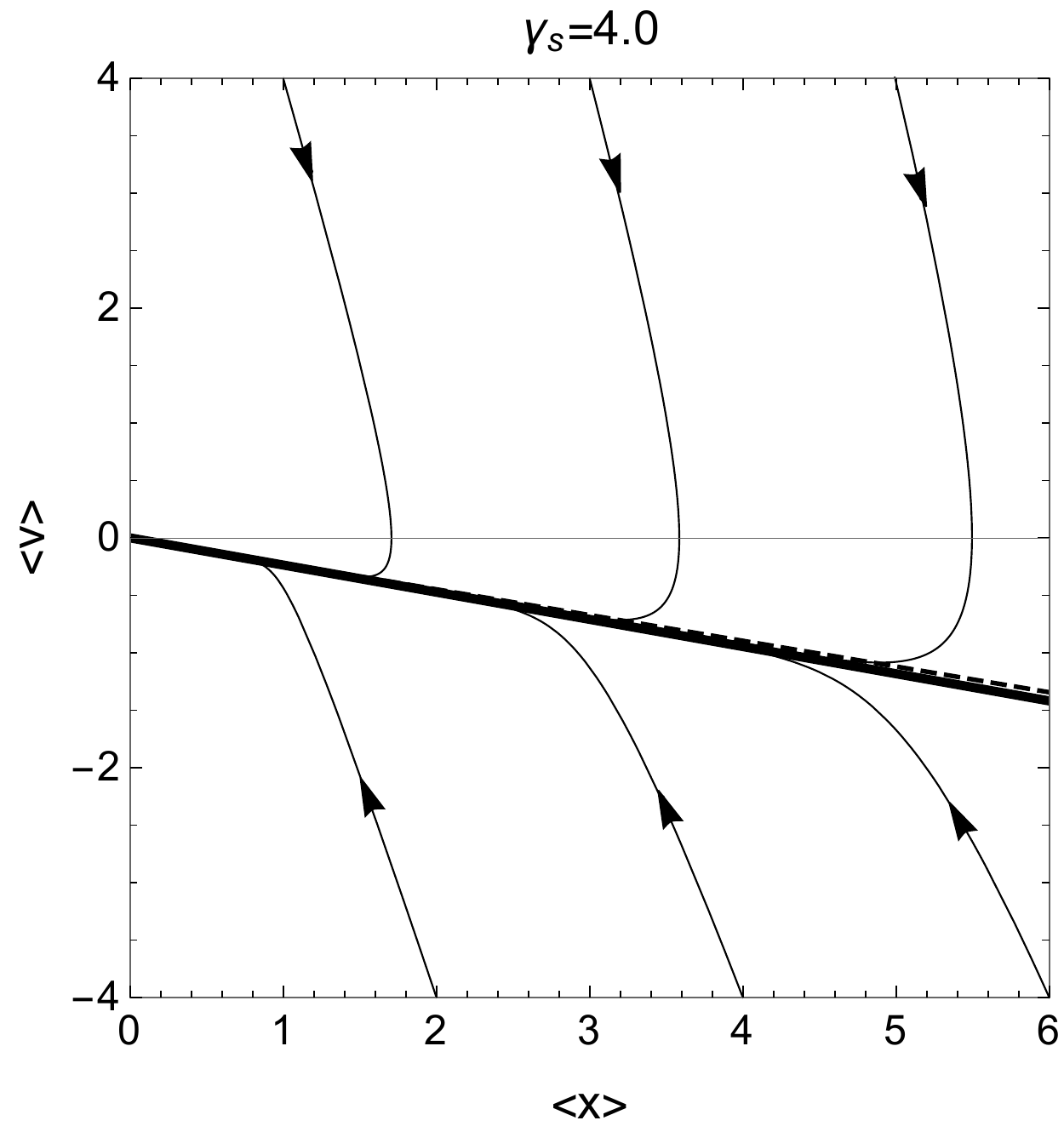}                  
\end{center}
\caption{Behavior of the solutions of the ODE system \eqref{zM}, for different initial data, with $\omega_0=1$ and with $\gamma_s=0.04$ (left panel), $\gamma_s=1.5$ (central panel) and $\gamma_s=4$ (right panel). The tiny solid lines correspond to individual trajectories, the thick solid lines denote the eigenvector $\mathbf{u}_-$ of the matrix $\mathbf{M}$ and the dashed lines represent the solution obtained with the closure given in Eq. \eqref{1stb}.}\label{fig:fig1}
\end{figure}

\subsubsection{Exact summation of the Chapman-Enskog expansion}
\label{sec:CEexp}

The Chapman-Enskog scheme, introduced in Sec. \ref{sec:DIP}, can be adapted to the reduction of the Brownian oscillator model as follows. 
The procedure starts from averaging Eq. \eqref{eq1} over noise,
\begin{equation}
  \langle \dot{\mathbf{x}}(t)\rangle = -\mathbf{M}\ \langle \mathbf{x}(t) \rangle \label{zM} \; .
\end{equation}
Solutions of the ODE system \eqref{zM} are portrayed in Fig. \ref{fig:fig1}.
One regards $\langle x(t)\rangle$ as the variable characterizing the reduced description, and assumes that the evolution of the fast variable $\langle v(t)\rangle$, after the initial layer, reaches a neighborhood of the slow manifold parameterized by $\langle x(t)\rangle$. 
Next, the variable $\langle v(t) \rangle$ is expanded in powers of $\varepsilon=\gamma^{-1}$, viz.
\begin{equation}
   \langle v(t) \rangle=\sum_{j=0}^{\infty}  \varepsilon^j  v^{(j)}(t) \, .
\label{expV}
\end{equation}
The coefficients $v^{(j)}(t)$ are found from the recurrence procedure
\begin{equation}
   v^{(j+1)} =-\sum_{k=0}^j 
    D_{CE}^{(k)}v^{(j-k)} \, , \, j\ge 1 \, ,\label{CEv}
\end{equation}
where the Chapman-Enskog operators $D_{CE}^{(k)}$ act on the coefficients $v^{(j)}$ as follows:
\begin{equation}
    D_{CE}^{(k)} v^{(j)}:= \frac{\partial   v^{(j)}}{\partial  \langle x \rangle} v^{(k)}  \, . \label{CEop}
\end{equation}
The recurrence equation \eqref{CEv} starts with $v^{(0)}=0$ and $v^{(1)}= -\omega_0^2 \langle x \rangle$.
A direct computation shows that the coefficients $v^{(j)}$ have the following structure to an arbitrary order $j\ge 0$:
\begin{equation}
    v^{(j)}(t)=- \bar{\alpha}_j \langle x(t) \rangle \label{closCE} \; ,
\end{equation}
with $\bar{\alpha}_{2j+1}>0$ and $\bar{\alpha}_{2 j}=0$.
Upon inserting the relation \eqref{closCE} into the recurrence equation \eqref{CEv}, the Chapman-Enskog method results in the nonlinear recurrence procedure for the coefficients $\bar{\alpha}_j$:
\begin{equation}
  \bar{\alpha}_{j+1}=\sum_{k=0}^j \bar{\alpha}_k \bar{\alpha}_{j-k} \quad , \quad j\ge 1 \, ,
  \label{CE}
\end{equation}
with the initial conditions $\bar{\alpha}_0=0$ and $\bar{\alpha}_1=\omega_0^2$. 
We observe that the term
\begin{equation}
    \alpha_1=\frac{\omega_0^2}{\gamma}  \, ,\label{1st}
\end{equation} 
recovers the drift coefficient in the Smoluchowski equation \eqref{smoluc}. 
The corresponding closure
\begin{equation}
    \langle v(t) \rangle = - \alpha_1 \langle x(t)\rangle \label{1stb}
\end{equation}
does not allow to accurately reconstruct the behavior of the trajectories of the dynamics \eqref{zM} in presence of moderate damping effects, as visible in the left and central panels of Fig. \ref{fig:fig1}.

We now aim at showing that the series
\begin{equation}
   \alpha=\sum_{j=0}^{\infty}\bar{\alpha}_j \varepsilon^j =\sum_{j=0}^{\infty} \alpha_j
\label{expA}
\end{equation}
can be summed up in closed form: this procedure will single out an algebraic invariant manifold for the linear ODE system \eqref{zM}.
We start by multiplying both sides of \eqref{CE} by $\varepsilon^{j+1}$ and then sum in $j$ from $1$ to $\infty$. We obtain:
\begin{equation}
 \varepsilon^{-1}\left[ \sum_{j=0}^{\infty} \bar{\alpha}_j \varepsilon^j - \bar{\alpha}_0-\bar{\alpha}_1\varepsilon \right]= \sum_{j=0}^{\infty}\varepsilon^j \left(\sum_{k=0}^{j}\bar{\alpha}_k \bar{\alpha}_{j-k}\right)-\bar{\alpha}_0^2, \label{sum1} 
\end{equation}
which, using \eqref{expA}, yields the invariance equation:
\begin{equation}
    \alpha^2-\gamma \alpha +\omega_0^2=0 \, . \label{inveq1}
\end{equation}
Note that Eq. \eqref{inveq1} is readily established from Eq. \eqref{IE2} by setting $\chi=\alpha/(m\omega_0^2)$ and $M=\beta m \omega_0^2 \langle x\rangle$.
The roots of the quadratic equation \eqref{inveq1} coincide with the two real-valued eigenvalues $\lambda_{\pm}$ of the drift matrix $\mathbf{M}$. 
Note that the eigenvalue $\lambda_+$ diverges in the limit $\varepsilon\rightarrow 0$.
Hence, since we look for bounded solutions to the invariance equation,
we set 
\begin{equation}
 \alpha=\lambda_- \label{drift} \; .
\end{equation} 
We remark that the parameter $\alpha$, in Eq. \eqref{drift}, corresponds to the exact summation of the Chapman-Enskog series \eqref{expA}, and it thus yields the desired correction of the drift term \eqref{1st} of the Smoluchowski equation up to an arbitrary order of $\varepsilon$. 

The same equation \eqref{inveq1} can also be derived, in a non-perturbative fashion, via the principle of Dynamic Invariance.
To this aim, we express the variable $\langle v (t)\rangle$ in terms of $\langle x (t)\rangle$ via the \textit{closure} $\Phi:\mathbb{R}\rightarrow\mathbb{R}$, which is endowed with the linear structure:
\begin{equation}
    \langle v(t)\rangle=\Phi[\langle x(t)\rangle]=-\alpha \langle x(t)\rangle, \label{exclos}
\end{equation}
where the parameter $\alpha>0$ depends on $\gamma$ and $\omega_0$. The relation \eqref{exclos} highlights a key aspect of the invariant manifold method: the variable $\langle v(t)\rangle$ depends on time only through the time dependence of the variable $\langle x(t)\rangle$.
Next, upon inserting the closure \eqref{exclos} in the ODE system \eqref{zM}, one obtains the so-called ``microscopic'' time derivative of $\langle v(t) \rangle$:
\begin{equation}
\partial_t^{(micro)}\langle v(t)\rangle:=\langle \dot{v}(t)\rangle=-\omega_0^2 \langle x(t) \rangle +\gamma \alpha \langle x(t) \rangle \, .
    \label{ce1}  
\end{equation}
We then define a projection operator $\mathcal{P}_x$, such that $\mathcal{P}_x\langle \dot{v}(t)\rangle|_{\langle v(t)\rangle=\Phi[\langle x(t)\rangle]}$ yields the evolution of the fast variable along the slow manifold parametrized by $\langle x(t) \rangle$. 
The action of $\mathcal{P}_x$ on $\langle \dot{v}(t)\rangle$ is expressed, in this case, via the chain rule:
\begin{equation}
    \mathcal{P}_x\langle \dot{v}(t)\rangle|_{\langle v(t)\rangle=\Phi[\langle x(t)\rangle]}=D_{\langle x(t) \rangle}\Phi[\langle x(t) \rangle]  \langle \dot{\mathbf{x}}(t)\rangle \, . \label{proj}
\end{equation}
In the sequel, to ease the notation, we shall denote the projected dynamics of $\langle{v(t)}$ by $\mathcal{P}_x\langle \dot{v}(t)\rangle$, where the closure \eqref{exclos} is implicitly assumed.
The ``macroscopic'' time derivative of $\langle v(t) \rangle$ is thus defined with the aid of the projection operator $\mathcal{P}_x$ as follows:
\begin{equation}
\partial_t^{(macro)}\langle v(t)\rangle :=\mathcal{P}_x\langle \dot{v}(t)\rangle
  =\alpha^2 \langle x(t) \rangle \, .
  \label{ce2} 
\end{equation}
The Dynamic Invariance principle, recall Eq. \eqref{InvEq1}, states that the two ``microscopic'' and ``macroscopic'' time derivatives \eqref{ce1} and \eqref{ce2} coincide, and the equality should hold independently of the value of the variable $\langle x(t) \rangle$: this leads again to the invariance equation \eqref{inveq1}.
It is worth pointing out that by reintroducing the expansion \eqref{expA} in Eq. \eqref{inveq1}, one may reconstruct ``backward''  the recurrence relation \eqref{CE} with the corresponding initial conditions. 
This observation clarifies that the invariance equation \eqref{inveq1} stands as the central result of the invariant manifold  method, whereas the Chapman-Enskog expansion can be interpreted an iterative procedure for solving the invariance equation via a power series representation. Relying on approximate solutions is, in fact, the only feasible approach when the invariance equation can not be solved analytically. Alternative iterative methods (based e.g. on the Newton's method), which may help circumvent some well-known instabilities appearing in low-order truncations of the Chapman-Enskog expansion, were considered in the framework of kinetic theory of gases \cite{colan07,colan07b}.
We conclude this Section by remarking that the invariant manifold method neglects, by construction, the fast relaxation dynamics in the initial layer, ruled by the eigenvalue $\lambda_+$, whereas it accurately captures the evolution along the slow manifold, encoded by the eigenvalue $\lambda_-$. 
A meaningful application of the method thus requires that the parameter $\gamma_s$ be large enough to guarantee the existence of an appropriate separation of time scales \cite{Rob08}. This, in turn, allows to retain the Markovian approximation also in the contracted description, as commonly done in the context of the Mori-Zwanzig projection operator approach \cite{Zwanzig}. 

\subsubsection{The Fluctuation-Dissipation theorem}
\label{sec:fdt}
Let us characterize, then, the fluctuations in Eq. \eqref{red}, by properly embedding the diffusion coefficient $\mathcal{D}_r$ in the framework of the Fluctuation-Dissipation theorem. 
By integrating Eq. \eqref{red} with a deterministic initial datum $x(0)=x$, one obtains:
\begin{equation}
 x(t)= e^{-\alpha t}x+\int_0^t e^{-\alpha (t-s)} dW(s)  \, . \label{solx}
\end{equation}
The two-time correlation function of the position variable can be then  calculated explicitly. It reads:
\begin{eqnarray}
    \langle x(s) x(t)\rangle&=& e^{-\alpha(t+s)}x^2 + 2 \mathcal{D}_r \int_0^{\min (s,t)} e^{-\alpha(t+s-2\tau)} d\tau \nonumber\\
    &=& \left(x^2 -\frac{\mathcal{D}_r}{\alpha}\right)e^{-\alpha(t+s)}+\frac{\mathcal{D}_r}{\alpha} e^{-\alpha|t-s|}\, .\label{d3}
\end{eqnarray}
We set $s=t$ and require that the stationary value of $\langle x(t)^2\rangle$ fulfills the Equipartition Theorem, namely:
\begin{equation}
    \lim_{t\rightarrow\infty} \langle x(t)^2\rangle= 
    (\beta m \omega_0^2)^{-1} \, . \label{d4}
\end{equation}
As a direct consequence, we  obtain a relation establishing a connection between the \textit{exact} drift coefficient $\alpha$ and the reduced diffusion coefficient $\mathcal{D}_r$:
\begin{equation}
    \alpha=\beta m \omega_0^2 \mathcal{D}_r \, . \label{einst2}
\end{equation}
Eq. \eqref{einst2} is an instance of the Fluctuation-Dissipation theorem of the II kind \cite{Kubo} for the reduced dynamics \eqref{red}. In fact, since for the harmonically bound particle the mobility takes the form $\chi=\alpha/(m\omega_0^2)$, Eq. \eqref{einst2} establishes the Einstein relation $\chi=\beta \mathcal{D}_r$, expressing the connection between the mobility and the diffusion coefficient of the reduced dynamics.

We also note that using Eqs. \eqref{1st} and \eqref{einst2}, it is possible to relate $\mathcal{D}_r$ to the diffusion coefficient $D$ of the original dynamics:
\begin{equation}
 \mathcal{D}_r=(\alpha_1)^{-1} \alpha D \, , \label{drd}   
\end{equation}
which offers a multi-level characterization of the fluctuations in the Brownian oscillator model.
One may expand $D_r$ in a power series in  $\varepsilon$, viz.:
\begin{equation}
    \mathcal{D}_r=\sum_{j=0}^{\infty} \overline{\mathcal{D}}_j\varepsilon^{j-1}=\sum_{j=0}^{\infty} \mathcal{D}_j \, .   \label{expD}
\end{equation}
Upon inserting \eqref{expA} and \eqref{expD} into \eqref{drd}, one obtains a hierarchy of equations relating, for each $j\ge 0$, the coefficients $\overline{\mathcal{D}}_j$ to the coefficients $\bar{\alpha}_j$ in \eqref{expA}:
\begin{equation}
\overline{\mathcal{D}}_j=(\bar{\alpha}_1)^{-1}\bar{\alpha}_{j} D \quad , \quad j\ge 0. \label{drd3}
\end{equation}
Thus, the leading-order term in Eq. \eqref{drd3} corresponds to
\begin{equation}
    \mathcal{D}_1=D \,, \label{D1}
\end{equation} 
which recovers the diffusion coefficient in the Smoluchowski equation \eqref{smoluc}. 

\subsection{Quantitative control of the reduction error}
\label{sec:estim}

We  provide, here, some quantitative estimate of the error introduced by the application of the reduction method to the Brownian oscillator model. There are two relevant sources of error coming with the proposed scheme. The first source, which is somehow intrinsic in the procedure, traces back to the moment parameterization introduced in Sec. \ref{sec:DIP} and is connected to the existence of an invariant manifold parameterized by the values of the slow variable. A proper choice of the initial data allows one to control such first contribution. A second source, instead, is related to the defect of invariance, and keeps track of the approximation introduced in solving the invariance equation. To see this, we denote by $\langle \mathbf{y}(t)\rangle=(\langle {y}(t)\rangle, \langle w(t)\rangle)$ the solutions of the ODE system \eqref{zM} supplied with the closure \eqref{exclos}, viz.:
\begin{equation}
 \langle w(t) \rangle = - \alpha \langle y(t) \rangle \, , \label{yclos}
\end{equation}
with deterministic initial datum $y(0)=y$.
Our purpose here is to compare in a quantitative way $\langle y(t)\rangle$ with $\langle x(t)\rangle$, the latter being the solution of the same ODE system when no closure is invoked. We shall finally give a pointwise in time {\em a priori} representation of the reduction error.

To be specific, we introduce the error terms $e_1(t):=\langle x(t)\rangle-\langle y(t)\rangle$, $e_2(t):=\langle v(t)\rangle-\langle w(t)\rangle$, and $e_3(t) :=\langle \dot{v}(t)\rangle-\langle \dot{w}(t)\rangle$, as well as the defect of invariance: 
\begin{equation}
\Delta_y:=(1-P_y)\langle \dot{w}(t)\rangle \label{def} \; ,   
\end{equation} 
which, by virtue of the closure \eqref{yclos}, takes the form:
\begin{equation}
    \Delta_y=-(\omega_0^2-\alpha \gamma+\alpha^2)\langle y(t) \rangle \label{defect} \;.
\end{equation}
Since Eq. \eqref{yclos} gives $\langle \dot{w}(t)\rangle = \alpha^2 \langle y(t)\rangle$, using Eq. \eqref{defect}, we arrive at:
\begin{eqnarray}
    e_3(t)&=&
    -\omega_0^2 \langle x(t)\rangle -\gamma \langle v(t)\rangle  - \alpha^2 \langle y(t)\rangle \nonumber\\
    &=&-\omega_0^2 (\langle x(t)\rangle-\langle y(t)\rangle)-\gamma (\langle v(t)\rangle-\langle w(t)\rangle)+\Delta_y \; , \label{wait}
\end{eqnarray}
which can thus be rewritten as:
\begin{equation}
e_3(t)=-\omega_0^2 e_1(t)-\gamma e_2(t)+\Delta_y \; . \label{e3}
\end{equation}
Then, noticing that $e_3(t)=\dot{e}_2(t)=\ddot{e}_1(t)$, we may rewrite Eq. \eqref{e3} as:
\begin{equation}
    \ddot{e}_1(t)+\gamma \dot{e}_1(t)+\omega_0^2 e_1(t)=\Delta_y \; , \label{e1final}
\end{equation}
in which $\Delta_y$ stands as the source term in the second-order linear ODE describing the dynamics of $e_1(t)$.

\begin{figure}[h!]
\begin{center}
         \includegraphics[width=0.32\textwidth]{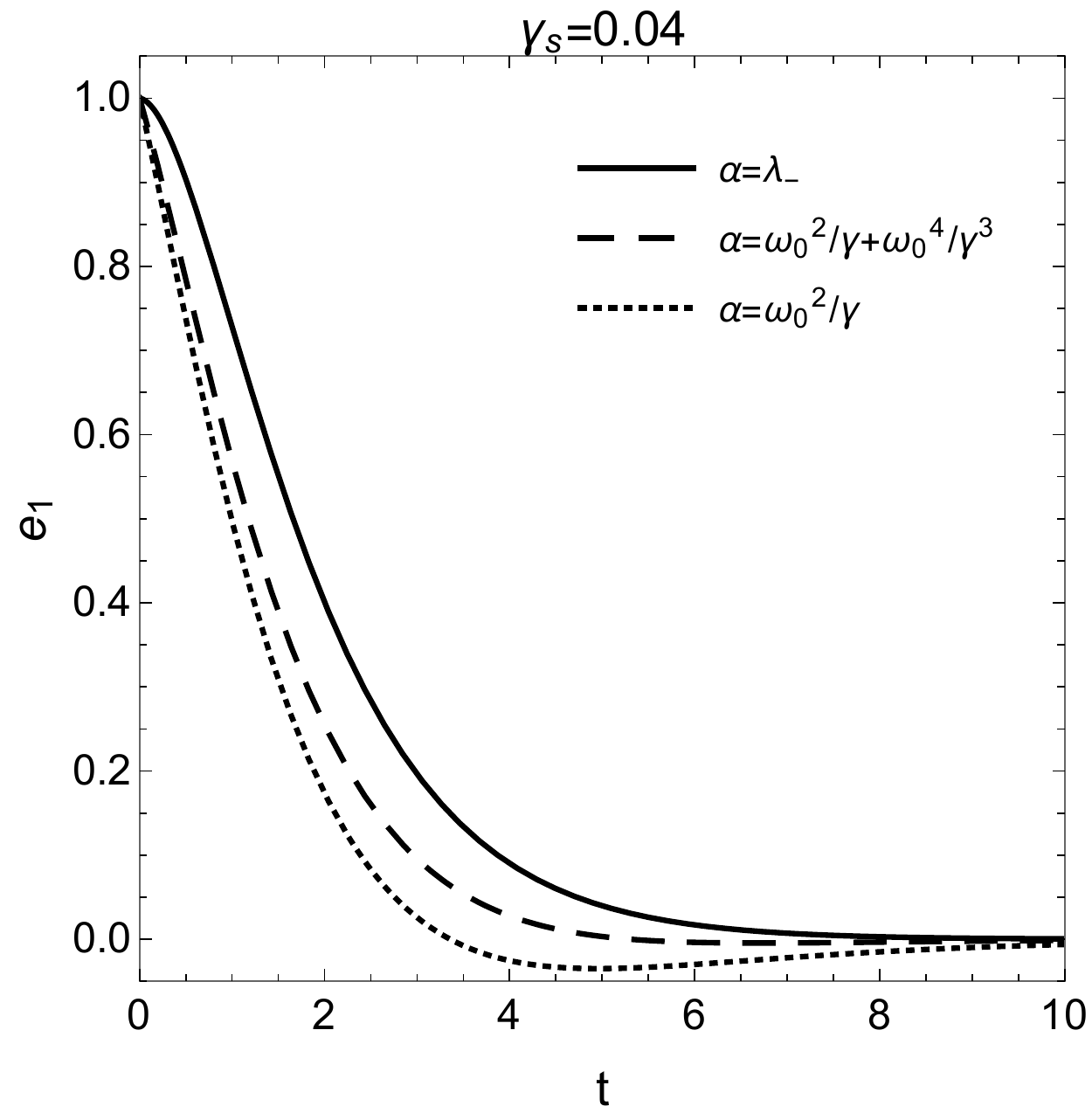}
         \includegraphics[width=0.32\textwidth]{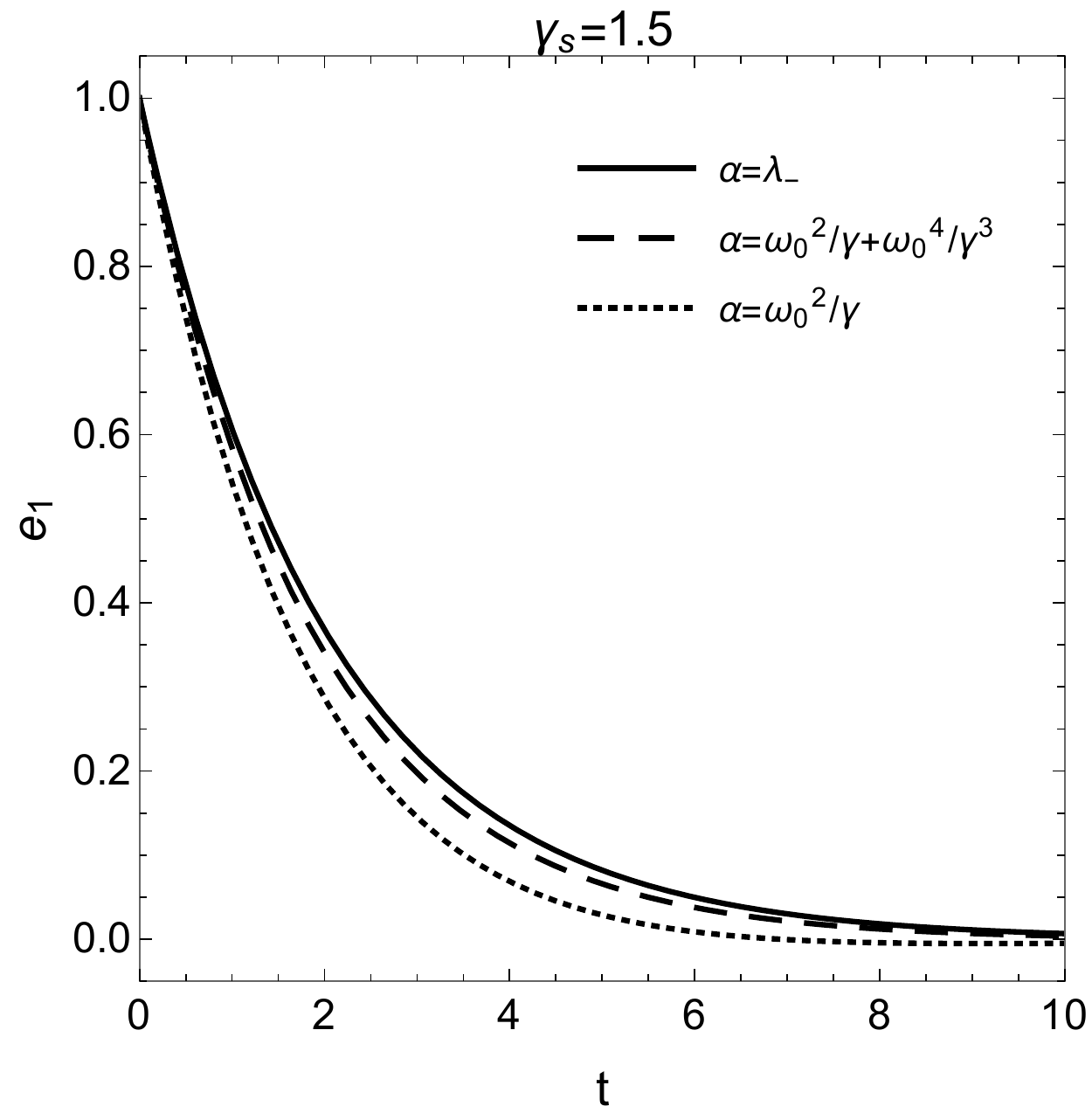}
         \includegraphics[width=0.32\textwidth]{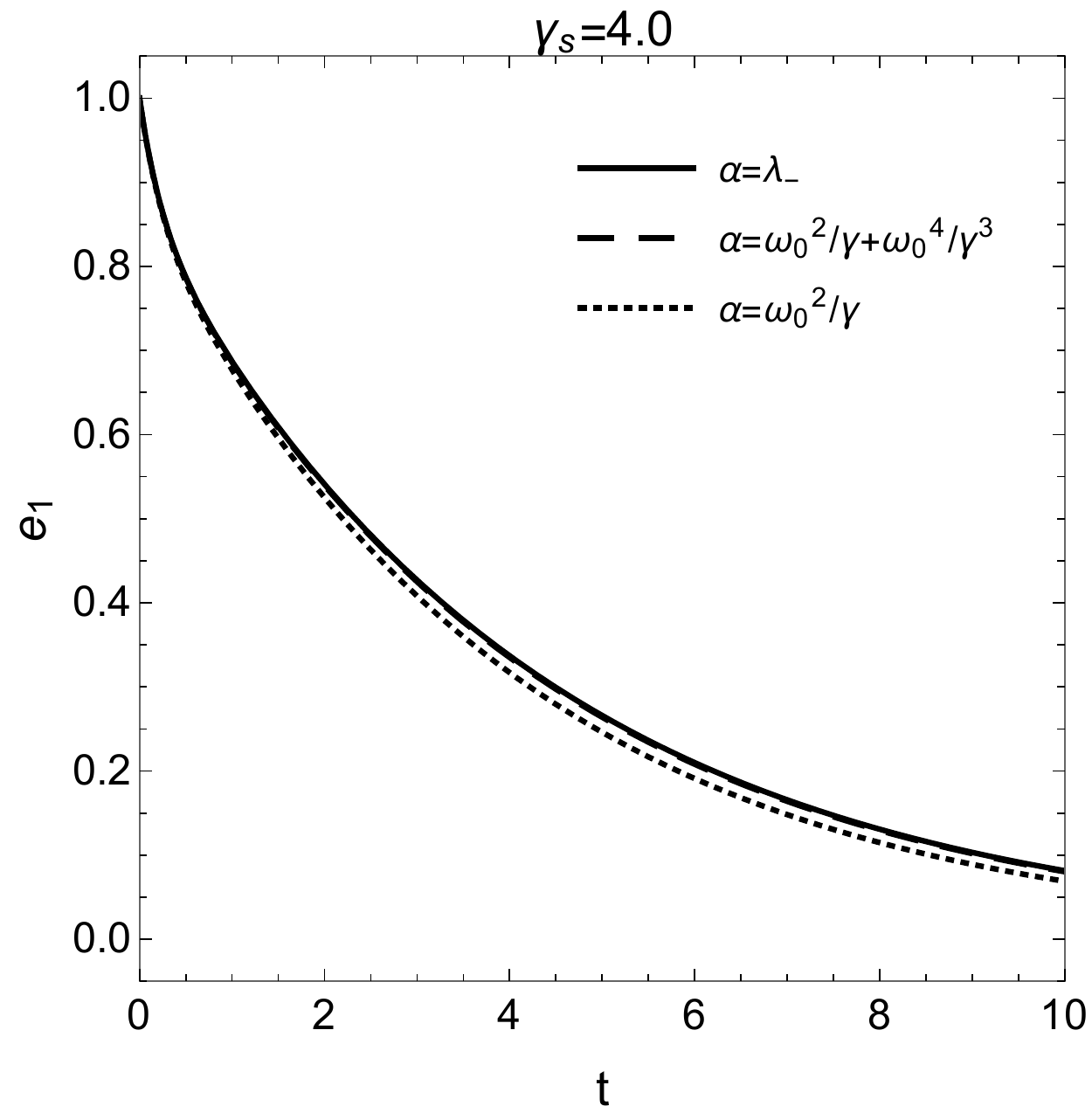}                  
\end{center}
\caption{Solutions of the ODE \eqref{e1final}, with $x=2$, $v=-1$, $y=1$ and with different choices of the closure \eqref{yclos}. Shown are the reduction errors corresponding to the first-order term of the Chapman-Enskog expansion (dotted curve), the corresponding third-order approximation of the same expansion (dashed curve) and the solution of the invariance equation (solid curve). The values of $\omega_0$ and $\gamma_s$ in the various panels correspond to those considered in Fig. \ref{fig:fig1}.}\label{fig:fig2}
\end{figure}

A double integration over time of Eq. \eqref{e1final} yields:
\begin{equation}
    e_1(t)=e_1(0)+e_2(0)t-\int_0^t ds \int_0^s d\tau (\omega_0^2 e_1(\tau)+\gamma e_2(\tau))+ \int_0^t ds \int_0^s d\tau \Delta_y \label{e1int} \; ,
\end{equation}
which can be cast in the form:
\begin{equation}\label{error-representation}
   e_1(t)+\omega_0^2   \int_0^t ds \int_0^s d\tau  \left[e_1(\tau)\right]+ \gamma \int_0^t ds \int_0^s d\tau  \left[e_2(\tau)\right] = \mathcal{R}(t) \; ,
\end{equation}
where the residual 
\begin{equation}\label{est4}
\mathcal{R}(t):=e_1(0) + e_2(0)t+\int_0^t ds \int_0^s d\tau  \Delta_y
\end{equation}
quantifies the quality of the reduction method. 
Note that the definition of the residual in  Eq. (\ref{est4}) includes the defect of invariance $\Delta_y$ expressed by Eq. \eqref{defect}. If $\Delta_y$ vanishes, then controlling the error of the reduction method amounts to guessing the initial value $y$ such that the first two terms on the r.h.s. of \eqref{est4} are small. The behavior of $e_1(t)$ for different values of $\gamma_s$ and for different choices of the closure \eqref{yclos} is displayed in Fig. \ref{fig:fig2}. In each panel, solid, dotted and dashed lines correspond to the solution of the invariance equation $\Delta_y=0$ (see Eq. \eqref{drift}), the first-order term of Chapman-Enskog expansion (cf. Eq. \eqref{1st}) and the third-order approximation of the same expansion, respectively.

\subsection{Response and correlation functions}
\label{sec:LRT}

We now turn to the study of correlation functions, which constitute a useful test-bed to assess the range of applicability of the proposed reduced description of the Brownian oscillator model.
Following the basic tenets of linear response theory, correlation functions are connected to the response of the system to an external stimulus; we refer the reader to Ref. \cite{Vulp} for an exhaustive review on this subject and also to the concise theoretical guidelines provided in Appendix \ref{app:appC}.
We suppose that the system described by Eq. \eqref{red} is initially in equilibrium with a heat bath at inverse temperature $\beta$. 
The stationary distribution of the reduced dynamics \eqref{red} takes the form:
\begin{equation}
 \rho_0(x)=\sqrt{\frac{\beta m \omega_0^2}{2 \pi }}\exp\left\{-\frac{1}{2}\beta m \omega_0^2 x^2\right\}. \label{rhored}
\end{equation}
We then probe the dynamics \eqref{red} by adding on the right hand side, at time $t=0$, a small time-dependent perturbation $F(t)$.
The perturbation induces 
the following structure of the Fokker-Planck equation:
\begin{eqnarray}
\frac{\partial \rho(x,t)}{\partial t}&=&\left(\mathcal{L}^*+\mathcal{L}_{ext}^*\right) \rho(x,t) \label{FPred}\\
\rho(x,0)&=&\rho_0(x), \nonumber
\end{eqnarray}
where the operator $\mathcal{L}^*$ and $\mathcal{L}_{ext}^*$ acts on probability densities as follows:
\begin{eqnarray}
\mathcal{L}^*\rho(x,t) &=&\left(\alpha\frac{\partial}{\partial x} x + \mathcal{D}_r \frac{\partial^2}{\partial x^2}\right)\rho(x,t) \label{L0red} \\
\mathcal{L}_{ext}^*\rho(x,t)&=& - F(t) \frac{\partial}{\partial x}\rho(x,t) \label{L1red} \; .
\end{eqnarray}
To write the response formula, we introduce the observable $$B(x)=-\rho_0(x)^{-1}\partial_x \rho_0(x),$$
which takes here the form:
\begin{equation}
B(x)=\beta m \omega_0^2 x \,. \label{opBred}
\end{equation}
We then look at the response of the observable $A(x)=x$. To this aim, we shall denote by $\langle \dots \rangle_0$ the complete average taken with initial density $\rho_0$.
According to the basic guidelines of linear response theory, the response function $R_{x,x}(t)$, for $t>0$, attains the structure:
\begin{eqnarray}
R_{x,x}(t)&=& \int A(x(t)) B(x) \rho_{0}(x) dx \nonumber\\
&=& C(0)^{-1}C(t) \, ,\label{resp3}
\end{eqnarray}
where $C(t)=\langle x(t) x\rangle_0$ is the autocorrelation function of the position variable, and
\begin{equation}
C(0)=\langle x^2 \rangle_0=\frac{\mathcal{D}_r}{\alpha}=\left(\beta m \omega_0^2\right)^{-1} \, . \label{x0}
\end{equation}
We also note that the drift coefficient $\alpha$ is connected to the \textit{autocorrelation time} $\tau_{c}$, defined as:
\begin{equation}
    \tau_c= C(0)^{-1}\int_0^\infty C(t)~ dt =\alpha^{-1} \label{tc} \, .
\end{equation}
Starting from \eqref{red}, $C(t)$ is found to obey for any $t>0$ the equation:
\begin{equation}
\frac{d}{dt}C(t)+\alpha C(t)=0 \, , \label{mu1}
\end{equation}
with the initial condition fixed by Eq. \eqref{x0}. 
The connection between correlation and response functions can be further unveiled as follows.
By the Wiener-Khinchin Theorem \cite{Risken}, the spectral density $S(\omega)$ of a stationary random process $x(t)$ is equal to the Fourier transform of its autocorrelation function, i.e.:
\begin{equation}
S(\omega)=\frac{1}{2\pi}\int_{-\infty}^{+\infty} e^{-i \omega t} C(t) dt \, .\label{power}
\end{equation}
The dynamics described by Eq. \eqref{red}, admits a  \textit{dynamical mobility} \cite{Kubo}  (or \textit{generalized susceptibility}) $\mu(\omega)$ of the form:
\begin{equation}
    \mu(\omega)=\frac{1}{\alpha+i\omega} \, . \label{corr3}
\end{equation}
By multiplying both sides of Eq. \eqref{mu1} by the factor $e^{-i \omega t}$, and by integrating over time from $0$ to $+\infty$, an integration by parts gives:
\begin{equation}
 (\alpha+i \omega) \int_0^\infty e^{-i \omega t} C(t) dt - C(0)=0 \, ,\label{mu2}
\end{equation}
which, using Eq. \eqref{corr3}, leads to the following remarkable expression of the mobility:
\begin{equation}
\mu(\omega)=C(0)^{-1} \int_{0}^{+\infty} e^{-i \omega t} C(t) dt \, . \label{mu3}
\end{equation}
Owing to the fact that 
$C(t)=C(-t)$ is a real, symmetric function of time, 
we use the relation \eqref{mu3} to reshape Eq. \eqref{power} in the form:
\begin{equation}
 S(\omega)=\frac{C(0)}{\pi}\Re[\mu(\omega)] \, , \label{corr4}   
\end{equation}
where $\Re[\mu(\omega)]$ denotes the real part of the mobility $\mu(\omega)$.
The relation \eqref{corr4} is a classical version of the Fluctuation-Dissipation theorem of the I kind \cite{Kubo}, as it connects the response to an external stimulus, represented by the dynamical mobility, to the fluctuations spontaneously produced in the system described by Eq. \eqref{red}, encoded by the spectral density.
By now putting the explicit expressions \eqref{x0} and \eqref{corr3} in Eq. \eqref{corr4}, we recover the standard Lorentzian form of the spectral density of the reduced dynamics \eqref{red}:
\begin{equation}
S(\omega)=\frac{1}{\pi}\frac{\mathcal{D}_r}{\omega^2+\alpha^2} \, . \label{corr5}
\end{equation}
One may analogously repeat the foregoing derivation for the original dynamics of the Brownian oscillator, described by Eq. \eqref{eq1}, which constitutes an exactly solvable example \cite{Risken,Pavl}. The invariant density for the unperturbed dynamics has the explicit representation:
\begin{equation}
    \rho_{0}(x,v)=\frac{m \omega_0 \beta}{2 \pi}\exp\left\{-\frac{m\beta}{2}(v^2 +\omega_0^2 x^2)\right\}  \, .\label{rho}
\end{equation}
We probe Eq. \eqref{eq1} by adding a time-dependent term $F(t)$ in the dynamics of the position variable and check the response in the variable $x$ itself, as above. 
The perturbed operator $\mathcal{L}_{ext}$ in the Fokker-Planck equation reads: 
\begin{equation}
\mathcal{L}_{ext}\rho(x,v,t)=-F(t)\frac{\partial}{\partial x}\rho(x,v,t) \, , \label{orig1}
\end{equation}
and it holds:
\begin{equation}
   A(x)=x \quad , \quad B(x)=\beta m \omega_0^2 x \, . \label{orig2}
\end{equation}
Using the explicit expression of the element $\mathbf{G}_{xx}$ of the Green's matrix, see Appendix \ref{app:appA}, we end up with the following response formula for the original dynamics:
\begin{equation}
    \mathcal{R}_{x,x}(t)= \frac{\lambda_+ e^{-\lambda_- t}-\lambda_- e^{-\lambda_+ t}}{\gamma_s},  \label{resp1} 
\end{equation}
which inherits the contributions of both the ``fast'' and the ``slow'' time scales of the system, encoded by the eigenvalues $\lambda_{\pm}$ of the matrix $\textbf{M}$ in Eq. \eqref{M}. For any finite time $t>0$, it makes thus sense to compare the response formulae computed with both the reduced and the original dynamics, Eqs. \eqref{resp3} and \eqref{resp1}, respectively. Thus, from \eqref{resp3} and \eqref{mu1} one finds the following structure for the response function of the reduced dynamics:
\begin{equation}
R_{x,x}(t)=e^{-\alpha t} \quad , \quad t>0 \, . \label{resp4}
\end{equation} 
Recalling \eqref{drift}, one thus finds:
\begin{equation}
    \left|\mathcal{R}_{x,x}(t)-R_{x,x}(t)\right|
    \le \frac{\lambda_-}{\gamma_s}. \label{conv}
\end{equation}
Using the explicit dependence of  $\lambda_-$ and $\gamma_s$ on $\gamma$, one finds that $\lambda_-/\gamma_s \rightarrow 0$ as $\gamma\rightarrow +\infty$, which implies the uniform-in-time convergence of $R_{x,x}(t)$ to $\mathcal{R}_{x,x}(t)$. 

We emphasize that the upper bound  \eqref{conv} guarantees that the response function of the reduced dynamics converges to the response function of the original dynamics in the high friction limit, namely when the time scale separation, controlled by $\gamma_s$, grows. In this context, it is instructive to observe in Figure \ref{fig:fig1}  the first plot in comparison with the third plot. 

\section{Conclusions}
\label{sec:concl}

In this work we considered a classical problem of statistical mechanics, concerning the extraction of a reduced description for Brownian dynamics in a confining potential. Adiabatic elimination techniques were already introduced in the 1970s to derive the Smoluchowski equation from the Kramers equation in the high friction limit. Application of the Chapman-Enskog scheme to the Fokker-Planck equation paved the way to a systematic derivation of the Smoluchowski formula and its higher order corrections via an expansion in power of the inverse friction coefficient. The same procedure is traditionally exploited, in kinetic theory of gases, to obtain the Euler and the Navier-Stokes equations from the Boltzmann equation. Nevertheless, the method is also known to suffer from the onset of short wavelength instabilities, which violate the H-theorem. As evidenced by the study of different  kinetic models, the failure of the Chapman-Enskog expansion does not lie in the scheme itself, but in its truncation to lower order levels.
The invariant manifold method is a non-perturbative reduction technique leading to hydrodynamic equations that are, instead, stable at all wavelengths. 
The method stipulates a condition of dynamic invariance which, for the class of kinetic models known as linearized Grad's moment systems, was shown to yield the same result as the exact summation of the Chapman-Enskog expansion.
The aim of the present work is, hence, to outline the use of the invariant manifold set-up to the reduction of a class of Brownian dynamics. \\
The main results can be summarized as follows. \begin{enumerate}
\item By exploiting the Dynamic Invariance principle, we derived an equation of invariance whose solutions generalize the structure of the mobility coefficient beyond the overdamped limit, thus making the reduced description prone to real world applications. While analytical solutions to the invariance equation are available in the presence of power law potentials, in the simplest case of the Brownian oscillator model we also succeed to sum up exactly the Chapman-Enskog expansion.
\item We obtained a quantitative estimate of the error encoded in the reduction procedure, which can be controlled through a suitable choice of the initial data and by minimizing the defect of invariance. 
\item We used linear response theory to shed light on the response functions due to the original and the reduced dynamics. We proved the convergence of the two response functions in the high friction limit.
\end{enumerate}

We believe that the procedure outlined in this work can provide useful insights on a wider class of Brownian dynamics.  
Even when exact solutions of the invariance equation are not available, the combined use of an iterative scheme of solution of the equation and the Fluctuation-Dissipation relation may help unravel meaningful reduced descriptions.\\

\textbf{Acknowledgments}\\
MC thanks L. Rondoni (Turin Polytechnic, Italy) for many useful discussions. AM thanks H. Duong (Birmingham, UK) for his KAAS seminar on a related topic as well as for his constructive ideas related to Section \ref{sec:estim}.




\appendix

\section{The Brownian oscillator model} 
\label{app:appA}
The dynamics of the Brownian oscillator is described by a system of linear stochastic differential equations (SDEs) written in the It\^{o} form:
\begin{equation}
    d \mathbf{x}(t)=-\mathbf{M}~ \mathbf{x}(t)~ dt+\mathbf{g}~ dW(t) \ ,  \quad \label{eq1b}
\end{equation}
where $\mathbf{x}(t)=(x(t), v(t))$, $\mathbf{g}=\left(0,\sqrt{\frac{2 \gamma}{\beta m}}\right)$, $W(t)$ is the one-dimensional Wiener process, and $\mathbf{M}$ is the \textit{drift} matrix
\begin{equation}
    \mathbf{M}=\begin{pmatrix}
0 & -1 \\
\omega_0^2 & \gamma
\end{pmatrix} \, , \label{Mb}
\end{equation}
with $\omega_0=\sqrt{k/m}$ the natural frequency of the oscillator equipped with mass $m$ and elastic constant $k$.
The eigenvalues of $\mathbf{M}$ take the form:
\begin{equation}
    \lambda_{\pm}=\frac{\gamma\pm \gamma_s}{2}, \label{eigenv}
\end{equation}
with $\gamma_s=\sqrt{\gamma^2-4 \omega_0^2}$. 
Upon averaging over noise, Eq. \eqref{eq1b} reduces to the linear system of ODEs:
\begin{equation}
  \langle \dot{\mathbf{x}}(t)\rangle = -\mathbf{M}\ \langle \mathbf{x}(t) \rangle \label{zM2} \; ,
\end{equation}
whose solutions $\langle \mathbf{x}(t)\rangle= \mathbf{G}(t)\mathbf{x}$ can be obtained via the Green's matrix $\mathbf{G}(t)\in \mathbb{R}^{2\times 2}$, with elements:
\begin{eqnarray}
    \mathbf{G}_{xx}&=&\frac{\lambda_+ e^{-\lambda_- t}-\lambda_- e^{-\lambda_+ t}}{\Delta}  \quad , \quad \mathbf{G}_{xv}=\frac{ e^{-\lambda_- t}- e^{-\lambda_+ t}}{\Delta}  \, ,\nonumber\\
   \mathbf{G}_{vx}&=&\omega_0^2\frac{ e^{-\lambda_+ t}- e^{-\lambda_- t}}{\Delta}  \quad  , \quad \mathbf{G}_{vv}= \frac{\lambda_+ e^{-\lambda_+ t}-\lambda_- e^{-\lambda_- t}}{\Delta} \, , \label{a:Gmatr}
\end{eqnarray}
with $\Delta=\lambda_+-\lambda_-$.
Using the Einstein relation $D=(\beta m\gamma)^{-1}$, the diffusion matrix $\mathbf{D}\in \mathbb{R}^{2\times 2}$ takes the form:
\begin{equation}
\mathbf{D}=
    \begin{pmatrix}
    0 & 0 \\
    0 & 2 D \gamma^2
    \end{pmatrix} \, . \label{MD}
\end{equation}
Correspondingly, the elements of the covariance matrix $\boldsymbol{\sigma}(t)\in \mathbb{R}^{2\times 2}$ read:
\begin{eqnarray}
    \boldsymbol{\sigma}_{xx}(t)&=&\frac{1}{\beta m \omega_0^2}\left[ \frac{\lambda_+ +\lambda_-}{\lambda_+ \lambda_-}+\frac{4 (e^{-(\lambda_+ + \lambda_-)t}-1)}{\lambda_+ +\lambda_-} \right.\nonumber\\
    &-&\left.\frac{1}{\lambda_+}e^{-2 \lambda_+ t}-\frac{1}{\lambda_-}e^{-2 \lambda_- t}\right] \, , \nonumber\\
 \boldsymbol{\sigma}_{xv}(t)&=& \boldsymbol{\sigma}_{vx}(t)=\frac{1}{\beta m \omega_0^2}\left[e^{-\lambda_+ t}-e^{-\lambda_- t}\right]^2 \, ,\nonumber\\
  \boldsymbol{\sigma}_{vv}(t)&=&\frac{1}{\beta m \omega_0^2}\left[\lambda_+ + \lambda_-+ \frac{4 \lambda_+ \lambda_-}{\lambda_+ \lambda_-}(e^{-(\lambda_+ + \lambda_-)t}-1) \right.\nonumber\\
  &-&\left. \lambda_+ e^{-2 \lambda_+ t}-\lambda_- e^{-2 \lambda_- t}\right] \, .\nonumber 
\end{eqnarray}
From the foregoing expressions, in the limit $t \rightarrow \infty$ we obtain the statement of the Equipartition Theorem:
\begin{equation}
   \bar{\boldsymbol{\sigma}}_{xx}=(\beta m \omega_0^2)^{-1} \, , \quad
   \bar{\boldsymbol{\sigma}}_{vv}=(\beta m)^{-1} \, , \quad
  \bar{\boldsymbol{\sigma}}_{xv}=\bar{\boldsymbol{\sigma}}_{vx}=0 \, .\label{eqp1}
\end{equation}
The solutions of the ODE system \eqref{zM2} may be cast in the form:
\begin{equation}
    \langle\mathbf{x}(t)\rangle=c_+ e^{-\lambda_+ t}\mathbf{u}_+ + c_- e^{-\lambda_- t}\mathbf{u}_- \label{eigen}
\end{equation}
with $c_{\pm}=(\lambda_\mp x+v)/\gamma_s$, where $\mathbf{u}_+=(-1,\lambda_+)$ and $\mathbf{u}_-=(1,-\lambda_-)$ denote the eigenvectors of the matrix $\mathbf{M}$. The origin $(0,0)$ is a stable node and represents the only equilibrium point of the dynamics, with $\lambda_-$ playing the role of the leading eigenvalue. Namely, when the spectral gap 
\begin{equation}
\gamma_s = \lambda_+ - \lambda_-     \label{spectralgap}
\end{equation}
becomes large, the trajectories of the system \eqref{zM} undergo an increasingly fast relaxation along the direction of $\mathbf{u}_+$, while they eventually settle, on a longer time scale, along the direction of the ``slow'' eigenvector $\mathbf{u}_-$,.

The generator associated to the SDE \eqref{eq1b} is:
\begin{equation}
\mathcal{L} =v\frac{\partial}{\partial x}-\omega_0^2 x\frac{\partial}{\partial v} +\gamma\left(-v \frac{\partial}{\partial v} + \frac{1}{\beta m} \frac{\partial^2}{\partial v^2}\right) \, . \label{L0}  
\end{equation}
Let $\mathcal{L}^*$ denote the $L^2$-adjoint of the generator of the process and by $\rho_{0}(x,v)$ the corresponding invariant probability density, which satisfies the equation $\mathcal{L}^*\rho_{0}(x,v)=0.$ The density $\rho_{0}(x,v)$ has the explicit representation: 
\begin{equation}
    \rho_{0}(x,v)=\frac{m \omega_0 \beta}{2 \pi}\exp\left\{-\frac{m\beta}{2}(v^2 +\omega_0^2 x^2)\right\}  \, .\label{rho}
\end{equation}

\section{Linear response theory for stochastic dynamics} 
\label{app:appC}

We retain the notation of Appendix \ref{app:appA} and consider the It\^{o} SDE in $\mathbb{R}^d$:
\begin{equation}
        d \mathbf{x}(t)=\mathbf{h}(\mathbf{x}(t))dt+\mathbf{g}~ d\mathbf{W}(t)\, , \label{ito2}
\end{equation}
with drift $\mathbf{h}(\mathbf{x}(t))$ and diffusion matrix $\mathbf{D}=\mathbf{g}~\mathbf{g}^T$.
We suppose that the system is initially in equilibrium with a heat bath at the inverse temperature $\beta$, and is described by a stationary distribution $\rho_0(\mathbf{x})$. We probe the system by introducing, at time $t=0,$ a small time-dependent perturbation of the form $F(t) \mathbf{K}(\mathbf{x})$:
\begin{equation}
        d \mathbf{x}(t)=\mathbf{h}(\mathbf{x}(t))dt+F(t) \mathbf{K}(\mathbf{x})dt+\mathbf{g}~ d\mathbf{W}(t)\, . \label{itopert}
\end{equation}
 The Fokker-Planck equation (or \textit{forward Kolmogorov equation}) attains the modified structure:
\begin{eqnarray}
    \frac{\partial \rho(\mathbf{x},t)}{\partial t}&=&\left(\mathcal{L}^*+\mathcal{L}_{ext}^*\right) \rho(\mathbf{x},t) \, , \label{FP} \\
    \rho(\mathbf{x},0)&=&\rho_0(\mathbf{x}) \, ,\label{FP0}
\end{eqnarray}
where $\mathcal{L}^*$ denotes the Fokker-Planck operator of the unperturbed dynamics, whereas $\mathcal{L}_{ext}^*$ is the operator induced by the perturbation. 
The operators $\mathcal{L}^*$ and $\mathcal{L}_{ext}^*$ act on probability densities as follows:
\begin{eqnarray}
    \mathcal{L}^*\rho(\mathbf{x},t)&=&-[ \nabla \cdot (\mathbf{h}(\mathbf{x})\rho(\mathbf{x},t))]+ \frac{1}{2}D^2 : (\mathbf{D}\rho(\mathbf{x},t)) \, , \label{L0a} \\
    \mathcal{L}_{ext}^*\rho(\mathbf{x},t)&=&-F(t)[ \nabla \cdot (\mathbf{K}(\mathbf{x})\rho(\mathbf{x},t))]\, . \label{L1}
\end{eqnarray}

The condition \eqref{FP0} expresses the fact that the initial datum is drawn from the distribution $\rho_0(\mathbf{x})$, which is invariant for the unperturbed dynamics.
To first order in the perturbation, one may write:
$$\rho(\mathbf{x},t)\simeq \rho_0(\mathbf{x})+\rho_1(\mathbf{x},t) \, ,$$ 
with
\begin{equation}
    \rho_1(\mathbf{x},t)=\int_0^t e^{\mathcal{L}^*(t-s)}\mathcal{L}_{ext}^* \rho_0(\mathbf{x}) ds \, .\label{p1c}
\end{equation}
By letting $\Delta A (t)$ denote the deviation of the expected value of the observable $A(\mathbf{x})$, computed with respect to the density $\rho(\mathbf{x},t)$ from the expected value computed with respect to $\rho_0(\mathbf{x})$, in the linear regime one thus finds:
\begin{equation}
\Delta A (t)=  \int A(\mathbf{x}) \rho_1(\mathbf{x},t) d\mathbf{x} \, , \label{intp}
\end{equation}
By inserting \eqref{p1c} in \eqref{intp}, one hence arrives to the linear response formula:
\begin{equation}
 \Delta A(t)=\int_0^t R_{A,B}(t-s) F(s) ds \, , \label{resp}   
\end{equation}
where $R_{A,B}$, called the \textit{response function}, describes the response of the observable $A(\mathbf{x})$ to the perturbation acting on the observable $B(\mathbf{x})$. With the aid of the generator $\mathcal{L}$ of the unperturbed process,
one may cast the expression of $R_{A,B}$ in the form of a two-time equilibrium correlation function, viz:
\begin{equation}
    R_{A,B}(t)=\int A(\mathbf{x}(t)) B(\mathbf{x}) \rho_{0}(\mathbf{x}) d\mathbf{x}\, , \label{resp2}
\end{equation}
where $B(\mathbf{x})$ takes the structure:
\begin{equation}
B(\mathbf{x})=-\left[\rho_{0}(\mathbf{x})^{-1}\nabla \cdot (\mathbf{K}(\mathbf{x})\rho_0(\mathbf{x}))\right] \, . \label{opB}
\end{equation}


\bibliographystyle{plain}
\bibliography{biblio}

\end{document}